# LONG-RANGE BEAM–BEAM EFFECTS IN THE TEVATRON*

V. Shiltsev[#], A. Valishev, FNAL, Batavia, IL, USA

*Abstract*

Long-range beam–beam effects occurred in the Tevatron at all stages (injection, ramp, squeeze, and collisions) and affected both proton and antiproton beams. They resulted in beam losses and emittance blow-ups, which occurred in remarkable bunch-to-bunch dependent patterns. On the way to record-high luminosities of the collider, many issues related to the long-range beam–beam interactions have been addressed. Below we present a short overview of the long-range beam–beam effects in the Tevatron. (For a detailed discussion on the beam–beam effects in the Tevatron please see reviews in Refs. [1–3] and references therein).

## HELICAL ORBITS IN TEVATRON

Beam–beam interactions in the Tevatron differ between the injection and collision stages. The helical orbits were introduced to provide sufficient separation between the proton and antiproton beams in order to reduce detrimental beam–beam effects, e.g. tune shifts, coupling, and high-order resonance driving terms. In $36 \times 36$ bunch operation, each bunch experienced 72 long-range interactions per revolution at injection, but at collision there were 70 long-range interactions and two head-on collisions per bunch at the Collider Detector at Fermilab (CDF) and D0 detectors (see Fig. 1). At a bunch spacing of 396 s$^{-9}$, the distance between the neighbour interaction points (IPs) was 59 m. In total, there were 138 locations around the ring where beam–beam interactions occurred. The sequence of 72 interactions out of the 138 possible ones differed for each bunch, hence the effects varied from bunch to bunch. The locations of these interactions and the beam separations change from injection to collision because of the antiproton cogging (relative timing between antiprotons and protons).

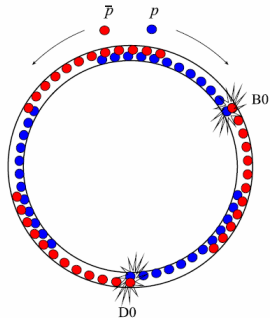

Figure 1: Schematic of proton (blue) and antiproton (red) bunches in the Tevatron and the two head-on collision locations B0 and D0.

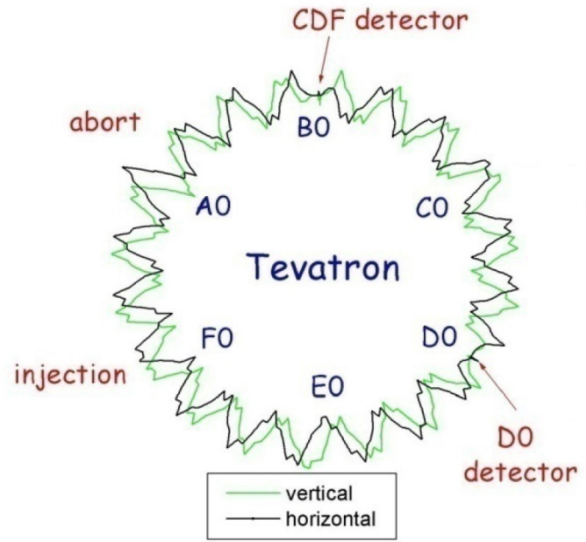

Figure 2: The pattern of the Tevatron helical orbits at the collision stage.

Initially, there were six separator groups (three horizontal and three vertical) in the arcs between the two main interaction points, B0 (CDF) and D0. During collisions, these separators form closed 3-bumps in each plane (see Fig. 2). However, the condition of orbit closure prevented running the separators at maximum voltages with the exception of horizontal separators in the short arc from B0 to D0. This limited separation at the nearest parasitic crossings 59 m away from the main IPs, aggravating the long-range beam–beam interaction. To increase separation at these parasitic crossings, three additional separators were installed to create closed 4-bumps both in the horizontal and vertical planes in the long arc (from D0 to B0) and in the vertical plane in the short arc. Each 3 m long HV separator (of which there were 24) was rated to operate with up to 300 kV over a 50 mm gap (horizontal/vertical) (see Fig. 3).

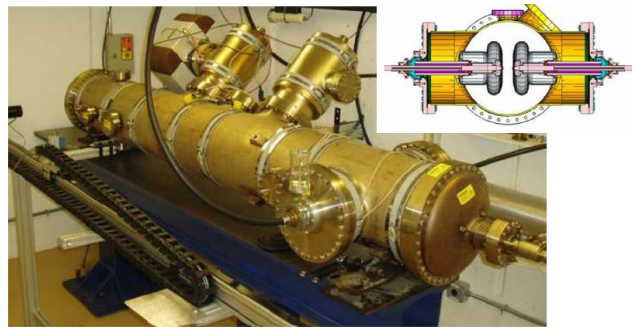

Figure 3: The Tevatron electrostatic HV separator.

There was some flexibility in the helix design for the preceding stages (injection, ramp, and squeeze). There were still some difficulties at these stages, including:

i) irregularities in betatron phase advance over the straight sections, especially A0;
ii) aperture restrictions (physical as well as dynamic) that limit the helix amplitude at injection and at the beginning of the ramp (see Fig. 4);

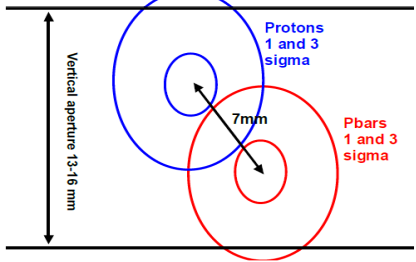

Figure 4: Schematic representation of one of the smallest separation locations at the C0 region inside the 16 mm aperture magnets. Long-range interaction at the spot caused significant beam losses and the small aperture magnets were taken out and replaced with 40 mm aperture dipoles in 2003.

iii) the maximum separator gradient of 48 kV/cm (limited by the separator spark rate) leads to a faster drop in separation, $d \sim 1/E$, than in the beam size, $\sigma \sim 1/E^{1/2}$, during the second part of the ramp above an energy of $E = 500$ GeV;
iv) the polarity reversal of the horizontal separation during the squeeze (to satisfy the needs of high energy physics (HEP) experiments) that leads to a short partial collapse of the helix.

Helical orbits were optimized many times over the course of Collider Run II in order to improve the performance of the machine. Our experience has shown that less than $S \sim 6\,\sigma$ separation resulted in unsatisfactory losses. Figure 5 shows the minimum radial separation $S$ during the ramp and squeeze with the initial helix design (blue, ca. January 2002) and an improved helix (red, ca. August 2004). The long-range interactions contribute a tune spread [1] of about:

$$\Delta Q \approx \sum_{parasitic\ encounters} \frac{2\xi}{S^2} \approx 0.008 \qquad (1)$$

as well as several units of chromaticity [4, 5]. For comparison, the head-on beam–beam tune shift parameters for both protons and antiprotons were about:

$$\xi = N_{IP} \frac{N_p r_p}{4\pi\varepsilon} \approx 0.018 - 0.025 \qquad (2)$$

where $r_p$ denotes the classical proton radius, $N_p$ and $\varepsilon$ are the opposite bunch intensity and emittance, respectively, and $N_{IP} = 2$ is the total number of head-on collisions per turn.

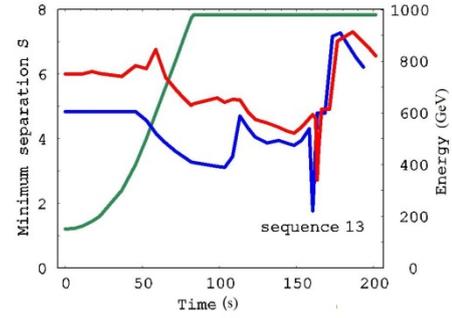

Figure 5: Minimum radial separation, Eq. (3), on ramp and during the low-beta squeeze. The green line represents the beam energy on the ramp. The blue and red lines represent $S(t)$ for the helix configurations used ca. January 2002 and August 2004, respectively (from Ref. [1]).

## BEAM–BEAM INDUCED LOSSES

As reported elsewhere, the beam–beam interactions had very detrimental effects on Collider performance early in Run II, but were eventually controlled via a number of improvements [1–3] (see Fig. 6).

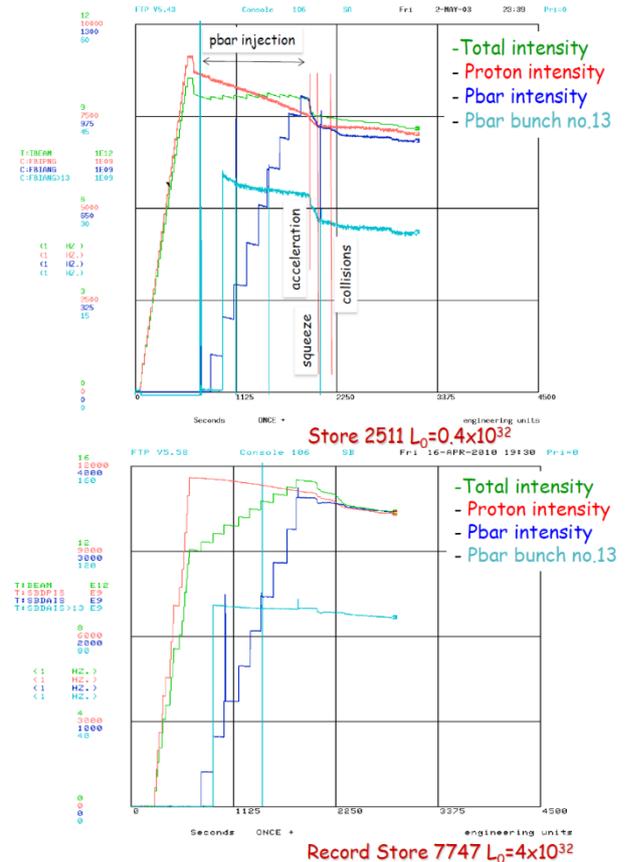

Figure 6: A typical plot of the collider 'shot' shows significant beam losses at all stages of the Tevatron cycle early in Run II (2003). A similar plot taken later in Run II shows greatly reduced inefficiencies and excellent performance in 2010.

Long-range beam-beam effects usually manifested themselves in reduction of beam lifetime and accelerated emittance growth. This accounted for as much as 50% luminosity loss early in Run II, down to ~10% loss at the end of the Run II. We observed no coherent effects that could be attributed to the LR beam–beam interactions.

At injection energy, LR beam-beam was the dominant factor for intensity losses both in proton and antiproton beams. This was especially noticeable for off-momentum particles, and strongly related to the tune chromaticity $Q'$ (strength of sextupoles). Figure 7 shows an interesting feature in the behaviour of two adjacent proton bunches (nos. 3 and 4). Spikes in the measured values are instrumental effects labelling the time when the beams are cogged (moved longitudinally with respect to each other). Initially, the bunches have approximately equal lifetimes. After injection of the second batch of antiprotons (four bunches each), the loss rate of bunch 4 greatly increased. After the first cogging, bunch 3 started to exhibit faster decay. Analysis of the collision patterns for these bunches allowed the pinpointing of a particular collision point responsible for the lifetime degradation [2].

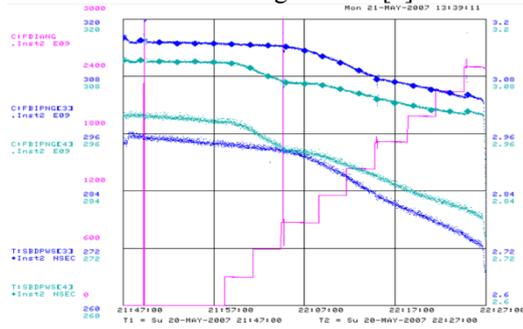

Figure 7: Intensity and rms length (s$^{-9}$) of proton bunches nos. 3 and 4 during injection of antiprotons (red line).

The particle losses for both beams on the separated orbits were larger at the higher intensities of the opposite beam (see Fig. 8) or, to be precise, larger at a higher brightness of the opposite beam (see Fig. 9), and were usually accompanied by longitudinal 'shaving' (preferential loss of particles with large momentum offset and corresponding reduction of the rms bunch length (see Fig. 10)).

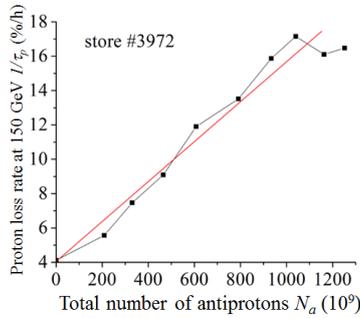

Figure 8: Proton loss rates at the energy of 150 GeV vs. the total number of injected antiprotons [1].

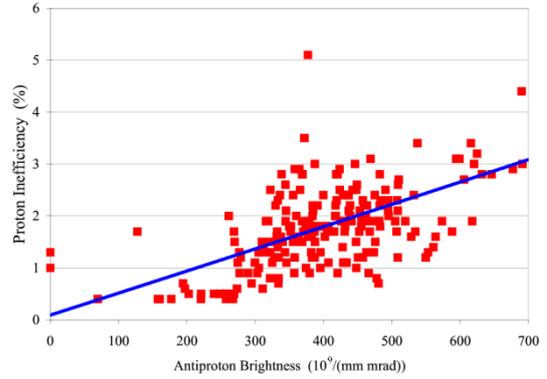

Figure 9: Proton losses on the energy ramp vs. antiproton brightness $N_a/\varepsilon_a$ [1].

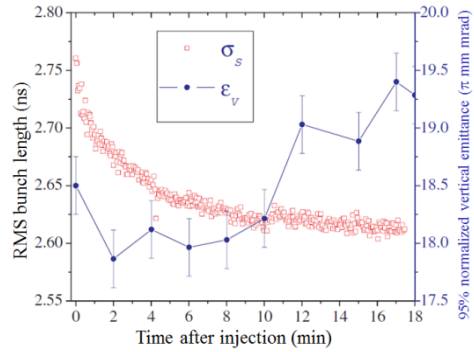

Figure 10: Time evolution of rms bunch length (red squares) and 95% normalized vertical emittance of antiproton bunch 1 (blue dots) after injection in store #3717 (8 August 2004). The error bars represent an rms systematic error in the flying wire emittance measurements [1].

The intensity decay was well approximated by [1]:

$$\frac{\Delta N_{a,p}}{N_{a,p}} = 1 - \frac{N(t)}{N(t=0)} \propto \sqrt{t} \cdot \varepsilon_{a,p}^2 \frac{N_{p,a}}{\varepsilon_{p,a}} Q'^2_{a,p} \cdot F(\varepsilon_L, Q_{x,y}, S_{a-p}) \quad (3)$$

The observed $\sqrt{t}$ dependence of beam intensity decay and bunch length is believed to be due to particle diffusion that leads to particle loss at physical or dynamic apertures (see Fig. 11). The major diffusion mechanisms are intrabeam scattering (IBS), scattering on the residual gas, and diffusion caused by RF phase noise. For example, if the available machine aperture is smaller than the beam size of the injected beam, the beam is clipped on the first turn, with an instantaneous particle loss. Such a clipping creates a step-like discontinuity at the boundary of the beam distribution that causes very fast particle loss due to diffusion. The diffusion wave propagates inward, so that the effective distance is proportional to $\sqrt{t}$. Consequently, the particle loss is also proportional to $\sqrt{t}$. To estimate such a 'worst-case loss', consider an initially uniform beam distribution: $f(I) = f_0 \equiv 1/I_0$, where $I_0$ is the action at the boundary. For sufficiently small time:

$t \ll I_0/D$, where $D$ is the diffusion coefficient, the diffusion can be considered one-dimensional in the vicinity of the beam boundary. Solving the diffusion equation:

$$\frac{\partial f}{\partial t} = D \frac{\partial}{\partial I}\left(I \frac{\partial f}{\partial I}\right) \quad (4)$$

gives the result:

$$f(I,t) = \frac{2f_0}{\sqrt{\pi}} \int_0^{(I_0-I)/\sqrt{4I_0 Dt}} e^{-\xi^2} d\xi \quad (5)$$

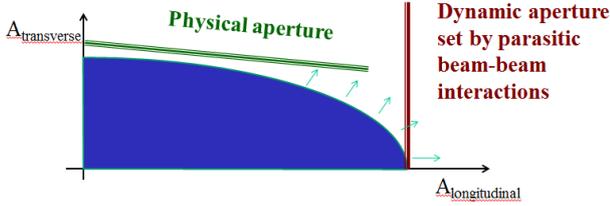

Figure 11: Schematic representation of the loss mechanism due to diffusion onto dynamic aperture set by the LR beam–beam interactions in the longitudinal–transverse action plane.

By integrating it over $I$, one obtains the dependence of particle population on time:

$$\frac{N(t)}{N_0} \approx 1 - \sqrt{\frac{t}{\tau}}, \quad \tau = \frac{\pi I_0}{4D}, \quad t \ll \tau \quad (6)$$

In the transverse degree of freedom, the Tevatron acceptance at 150 GeV on the helical orbit is about $I^{tr}_0 \approx 8{-}13$ π mm mrad, depending on the pre-shot machine tune-up, while the emittance growth rate is about $D^{tr} \approx 0.15{-}0.25$ π mm mrad/h, chiefly from external noises and scattering on the residual gas. From Eq. (6), one can obtain a lifetime of $\tau \approx 30{-}80$ h. In addition, diffusion in the longitudinal plane with a rate $D^{long} \approx 0.03{-}0.3$ rad$^2$/h can lead to lifetimes of $\tau \approx 10{-}100$ h in the case where the longitudinal aperture is limited only by the RF bucket size $\sqrt{I_0^{long}} \approx 2$ rad. Not all the numbers used above are well known, but we believe they are in the indicated ranges.

In reality, the machine acceptance is determined by the interplay between the physical and dynamic apertures. The latter is a strong function of the synchrotron action, and beam–beam interactions drastically reduce the dynamic aperture for synchrotron oscillation amplitudes close to the bucket size. Naturally, such an aperture reduction is stronger for larger values of chromaticity.

Notably, the proton inefficiencies were higher than the antiproton ones, despite a factor of 3–5 higher proton intensity. That was due to significantly smaller antiproton emittances (see Eq. (3) above).

During low-beta squeeze the beams briefly (for ~2 s) came within 2–2.5 σ at 1 parasitic IP. This caused sharp loss spikes. In general, the beam intensity losses were dependent on:

i) the chromaticities $Q'_{x,y}$, and special measures were taken for their reduction (reduction of impedance and implementation of octupoles and feedback systems allowed $Q'$ to decrease to almost zero);

ii) beam separation:

$$S = \sqrt{(\Delta x/\sigma_{x\beta})^2 + (\Delta y/\sigma_{y\beta})^2} \quad (7)$$

e.g. at collisions there were four crossings at 5.8–6 σ separation that were essential, the remaining LR's were at 8–10 σ;

iii) during the colliding beams stores—complex interplay between the head-on and parasitic long-range interactions (the head-on tune shifts up to about $\xi = 0.020{-}0.025$ for both protons and antiprotons, in addition to the long-range tune shifts of $\Delta Q^p = 0.003$ and $\Delta Q^a = 0.006$, respectively, see Ref. [3]);

iv) on the second order betatron tune chromaticity $Q'' = d^2Q/d(\Delta p/p)^2$ (numerical modelling [2] indicated, and it was later confirmed by experiments that the deterioration of the proton life time was caused by a decrease of the dynamical aperture for off-momentum particles at high $Q''$);

v) on the bunch position in the train (there were remarkable differences in the dynamics of individual bunches—see below).

At the end of Run II, the antiproton intensity lifetime deterioration due to the beam–beam effects was much smaller than the proton one, and was found to scale approximately as [1]:

$$\left(\frac{1}{\tau_a}\right)_{BB} = \left(\frac{dN_a}{N_a dt}\right)_{BB} \propto N_p \frac{\varepsilon_a^2}{S^3}, \quad (8)$$

where $S$ stands for the beam–beam separation (helix size).

## PATTERNS OF BEAM–BEAM EFFECTS

All beam dynamics indicators were dependent on the bunch position in the train of bunches (there were three train of 12 bunches in each beam)—beam orbits and coupling (of about 40 microns (see Fig. 12)), tunes (by as much 0.005 as shown in Fig. 13) and chromaticities (up to six units (see Fig. 14)).

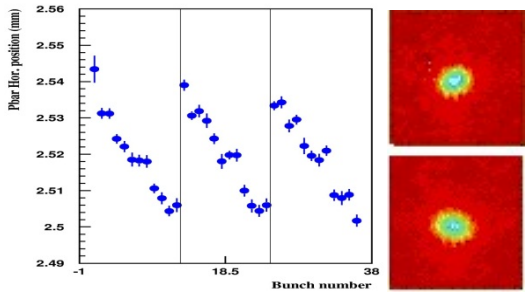

Figure 22: Antiproton horizontal orbit variations along the bunch train for comparison. The pbar rms horizontal betatron size at the location of the synchrotron light monitor [6] is equal to ~0.3 mm. 2D beam images on the right are for bunches #1 (top) and #8 (bottom). Different tilts of the images indicate a significant difference in local coupling.

Similar type differences (though smaller—proportional to the intensity of the opposite beam) took place for the proton bunches. The observed variations data are in good agreement with analytical calculations [1, 2, 5].

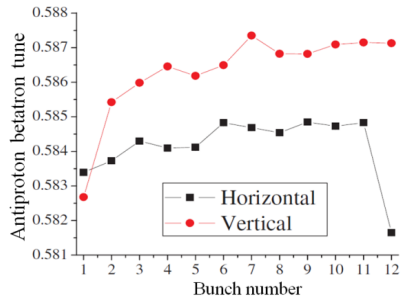

Figure 13: Horizontal and vertical antiproton tunes vs. bunch number in the bunch train measured by 1.7 GHz Schottky monitor [4] ~3 h into store #3678 (27 July 2004) [1].

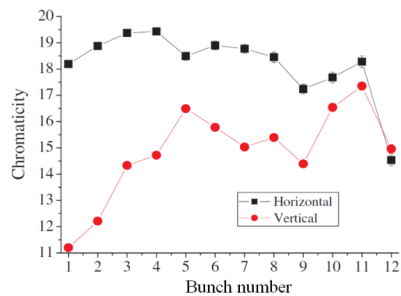

Figure 14: Antiproton chromaticities measured by the 1.7 GHz Schottky monitor vs. bunch number for store #3678 (27 to 28 July 2004) [1].

It is not surprising that with such significant differences in tunes and chromaticities, the antiproton and proton bunch intensity lifetime and emittance growth rates vary considerably from bunch to bunch. The orbit difference did not produce adverse effects on the performance. As an illustration, Fig. 15 shows the vertical emittance blow-up early in an HEP store for all three trains of antiproton bunches.

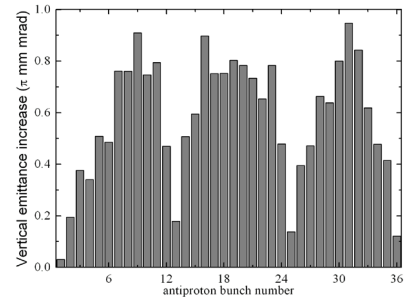

Figure 15: Antiproton bunch emittance increase over the first 10 minutes after initiating collisions for HEP store #3231 with an initial luminosity $L = 48 \times 10^{30}$ cm$^{-2}$ s$^{-1}$.

One can see a remarkable distribution along the bunch train, which gave rise to the term 'scallops' (three scallops in three trains of 12 bunches) for this phenomenon—the end bunches of each train exhibit lower emittance growth than the bunches in the middle of the train. Because of the three-fold symmetry of the proton loading, the antiproton emittance growth rates are the same within 5–20% for corresponding bunches in different trains (in other words, bunches #1, #13, and #25 have similar emittance growths). The effect is dependent on the antiproton tunes, particularly on how close each bunch is to some important resonances—in the case of the Tevatron working point, these are fifth-order (0.600), seventh-order (0.5714), and twelfth-order (0.583) resonances. For example, the scallops occur near the fifth-order resonances $nQ_x + mQ_y = 5$, such as $Q_{x,y} = 3/5 = 0.6$. Smaller but still definite scallops were also seen for protons if the proton tunes are not optimally set. After the initial 0.5–1 h of each store, the growth rate of each bunch decreased significantly. Various methods have been employed to minimize the development of scallops (including a successful attempt to compensate one bunch's emittance growth with a Tevatron electron lens (TEL), see Ref. [7]), but carefully optimizing the machine tunes was found to be the most effective, e.g. the vertical tune changes as small as −0.002 resulted in significant reduction of the amplitude of the scallops.

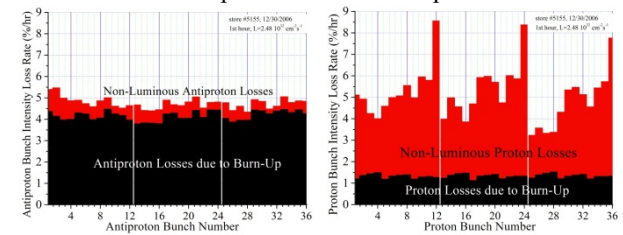

Figure 16: (a - left) proton-bunch intensity loss rates and (b - right) antiproton-bunch intensity loss rates at the beginning of Tevatron store #5155, 30 December 2006, with an initial luminosity $L = 250 \times 10^{30}$ cm$^{-2}$ s$^{-1}$ (from Ref. [7]).

The attrition rate of protons and antiprotons due to their interaction with the opposite beam varied bunch-by-bunch and is especially large at the beginning of the HEP stores where the total proton beam–beam tune shift parameter peaks. Figure 16(a) shows a typical distribution of proton loss rates $(dN_p/N_p)/dt$ at the beginning of a high-luminosity HEP store. Bunches #12, 24, and 36 at the end of each bunch train typically lost about 9% of their intensity per hour while other bunches lost only 4–6%/h. These losses were a very significant part of the total luminosity decay rate of about 20% per hour (again, at the beginning of the high luminosity HEP stores). The losses due to inelastic proton–antiproton interactions $dN_p/dt = \sigma_{int} L$ at the two main IPs ($\sigma_{int}$ = 0.07 barn) were small (1–1.5%/h) compared to the total losses. Losses due to inelastic interaction with the residual vacuum and due to leakage from the RF buckets were less than 0.3%/h. The single largest source of proton losses is the beam–beam interaction with the antiprotons. Such a conclusion is also supported by Fig. 16(a), which shows a large bunch-to-bunch variation in the proton loss rates within each bunch train, but very similar rates for equivalent bunches, e.g. bunches #12, 24, and 36. On the contrary, antiproton intensity losses $dN_a/dt$ were about the same for all of the bunches (see Fig. 16(b)) as they are mostly due to luminosity burn-up and not determined by beam–beam effects (the latter are labelled as a 'non-luminous' component of the loss rate).

The remarkable distribution of the proton losses seen in Fig. 16, e.g. the particularly high loss rates for bunches #12, 24, and 36, is usually thought to be linked to the distribution of betatron tunes along the bunch trains. Bunches at the end of the trains have their vertical tunes closer to the $7/12 \approx 0.583$ resonance lines and, therefore, have higher losses. The average Tevatron proton tune $Q_y$ of about 0.588–0.589 lies just above this resonance, and the bunches at the end of each train, whose vertical tunes are lower by $\Delta Q_y = -(0.002-0.003)$ due to the unique pattern of long-range interactions, are subject to stronger beam–beam effects. The tunes $Q_y$ $Q_x$ were carefully optimized by the operation crew to minimize the overall losses of intensity and luminosity. For example, an increase of the average vertical tune by quadrupole correctors is not possible because it usually results in higher losses and scallops as small amplitude particle tunes move dangerously close to the $3/5 = 0.600$ resonance. The Tevatron electron lenses did reduce by a factor of >2 the proton losses out of bunches #12, 24, and 36 (see Fig. 17) (for more details please refer to Refs. [3, 7, 8]).

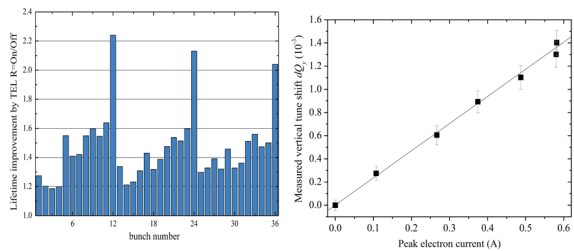

Figure 17: Proton bunch lifetime improvement factor due to (a - left) TEL), and (b - right) tuneshift vs. the TEL current [7].

## NOTE ON BEAM–BEAM SIMULATIONS

We would like to draw attention to the fact that for most of Collider Run II we had trustable numerical models and simulation tools for stored beam physics analysis and weak–strong beam–beam modelling, which were used to study the beam–beam effects in the Tevatron [2]. Our simulations correctly described many observed features of the beam dynamics, had predictive power, and have been particularly useful for supporting and planning changes of the machine configuration (see Figs. 18 and 19). We also had very practical computations of the resonant driving terms [9].

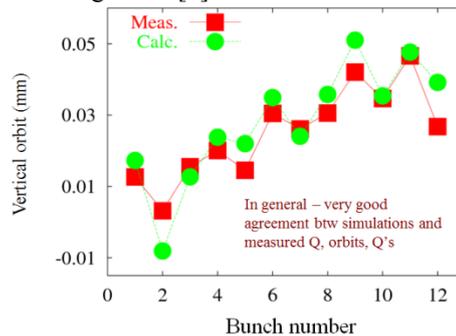

Figure 18: Bunch by bunch antiproton vertical orbits. Squares, measurements; circles, Lifetrac simulations [2].

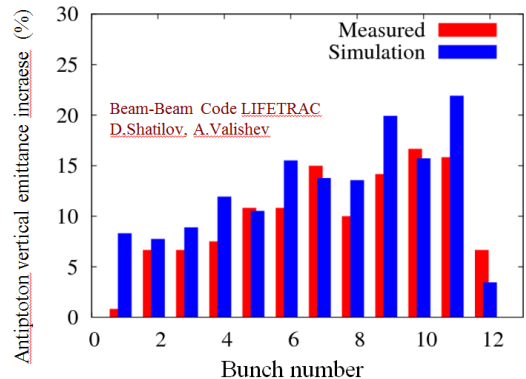

Figure 19: Bunch-by-bunch antiproton emittance growth. Measured in store 3554 (red) and simulated with Lifetrac (blue) [2].

## SUMMARY

Long-range beam–beam effects occurred in the Tevatron at all stages (injection, ramp, squeeze, and collisions) and in both beams. They resulted in beam losses and emittance blow-ups—with bunch-to-bunch dependent patterns. Careful optimization of helical orbit separation and many operational tune-ups and upgrades have led essentially to putting the effects upon the luminosity under control by the mid to end of Run II. Trustable weak–strong simulations had helped us a lot. Compensation of the LR beam–beam effects by TELs has been demonstrated.